\documentclass[prl,twocolumn,showpacs,epsfig,longeq]{revtex4}
\usepackage{graphicx}
\usepackage{dcolumn}
\usepackage{amssymb}
\usepackage{bm}
\usepackage{epsfig}
\usepackage{psfrag}
\RequirePackage[pdftex,pdfpagemode=none, pdftoolbar=true,
                pdffitwindow=true,pdfcenterwindow=true]{hyperref}

\def\3dots{\:\raisebox{-0.5ex}{$\stackrel{\textstyle.}{:}$}\:}
\def\beq{\begin{equation}}
\def\eeq{\end{equation}}
\def\bea{\begin{eqnarray}}
\def\eea{\end{eqnarray}}
\begin{document}

\title{Binding of Nucleobases with Single-Walled Carbon Nanotubes}

\author{Anindya Das,$^{1}$ A. K. Sood,$^{1,} $\thanks{Corresponding author: asood@physics.iisc.ernet.in}\\
Prabal K. Maiti,$^{2}$ Mili Das,$^{3}$ R. Varadarajan,$^{3}$ and C. N. R. Rao$^{4}$}

\affiliation{
$^{1}$Department of Physics, Indian Institute of Science, Bangalore 560012, India \\
$^{2}$Center for Condensed Matter Theory, Department of Physics, Indian Institute of Science, Bangalore 560012, India \\
$^{3}$Molecular Biophysics Unit, Indian Institute of Science, Bangalore-560012, India\\
$^{4}$Chemistry \& Physics of Materials Unit, Jawaharlal Nehru Centre for Advanced Scientific Research, Bangalore-560064, India
}
\begin{abstract}
We have calculated the binding energy of various nucleobases (guanine (G), adenine (A), thymine (T) and cytosine (C)) with (5,5) single-walled carbon nanotubes (SWNTs) using ab-initio Hartre-Fock method (HF) together with force field calculations. The gas phase binding energies follow the sequence G $>$ A $>$ T $>$ C. We show that main contribution to binding energy comes from van-der Wall (vdW) interaction between nanotube and nucleobases. We compare these results with the interaction of nucleobases with graphene. We show that the binding energy of bases with SWNTs is much lower than the graphene but the sequence remains same. When we include the effect of solvation energy (Poisson-Boltzman (PB) solver at HF level), the binding energy follow the sequence G $>$ T $>$ A $>$ C $>$, which explains the experiment\cite{zheng} that oligonucleotides made of thymine bases are more effective in dispersing the SWNT in aqueous solution as compared to poly (A) and poly (C). We also demonstrate experimentally that there is differential binding affinity
of nucleobases with the single-walled carbon nanotubes (SWNTs) by directly
measuring the binding strength using isothermal titration (micro) calorimetry.
The binding sequence of the nucleobases varies as thymine (T) $>$ adenine (A)
$>$ cytosine (C), in agreement with our calculation.
\end{abstract}

\maketitle

\section{Introduction}

Single-walled carbon nanotubes (SWNTs) are one-dimensional systems with different diameters and chiralities. They have large surface areas \cite{peigney,eswaramoorthy,bacsa} with the   electrons in intimate contact with the environment. The superb mechanical \cite{salvetat} and electrical \cite{colbert} properties of SWNT have potential for many applications such as nanoelectronics \cite{collins,tans}, actuators \cite{baughman}, chemical \cite{kong,staii}, and flow \cite{ghosh} sensors. Recently, there are potential applications of DNA coated CNTs as a biosensor \cite{lin,cassell}. Single-stranded DNA (ss-DNA) coated SWNTs field effect transistor (FET) has been used to detect various odors \cite{staii}, DNA hybridization \cite{star} and conformation changes in DNA \cite{heller} in presence of ionic concentration. It has been also reported that DNA can be inserted into CNTs, which can be further potential applications of this nano-bio system.
	
	During synthesis SWNTs appear in bundles, but for their potential application we need isolated SWNT. It is, therefore, a big challenge to disperse SWNT bundles in aqueous solution. It has been reported that single-walled carbon nanotube bundles are effectively dispersed in water on sonication in the presence of single stranded DNA (ssDNA) \cite{zheng}. The dispersity is dependent on the oligonucleotide sequence. Jagota $\textit{et al}$. \cite{zheng} has shown that poly (T) has higher efficiency in dispersion the nanotubes compared to  ploy(A) and ploy(C). Therefore, the wraping of carbon nanotubes by ss-DNA is sequence dependent. It was also found that the best separation was obtained with sequence of repeats alternating G and T (poly (GT)) \cite{mzheng}. Studies on the nature of interaction of ssDNA, dsDNA and oligonucleotides with nanotubes have made use of classical molecular dynamics simulations, besides experimental techniques such as ion exchange chromatography (IEC) \cite{mzheng}, atomic force microscopy (AFM), resonance Raman spectroscopy (RRS), photoluminescence (PL) \cite{chou}, linear dichroism (LD) \cite{rajendra} and directional optical absorbance \cite{meng}. There has, however, been no estimation of the strength of interaction of the individual nucleobases, adenine (A), cytosine (C), thymine (T) and guanine (G) with the SWNTs. The objective of this study is to determine the binding strength of the different nucleobases with SWNTs using ab-initio quantum chemical as well as classical force field calculations. We have also determined the binding strength of thymine and cytosine with the SWNTs by employing isothermal titration calorimetry (ITC), which has emerged as a powerful tool to study the thermodynamics of protein-protein \cite{thompson}, DNA-protein \cite{kunne} and protein-lipid interactions \cite{wenk}. The advantage of ITC over kinetic methods is that the same experiment provides the binding constant as well as the enthalpy of binding. In our study we have titrated aqueous SWNT solutions against aqueous solutions of nucleobases. Our results show that the binding affinity of thymine with SWNTs is higher than adenine and cytosine. We believe that the present study, besides providing valuable insight into the interactions of SWNTs with nucleobases, may be of potential use in nanotechnology.

\begin{figure}[tbp]
\begin{center}
\leavevmode
\includegraphics[width=0.475\textwidth]{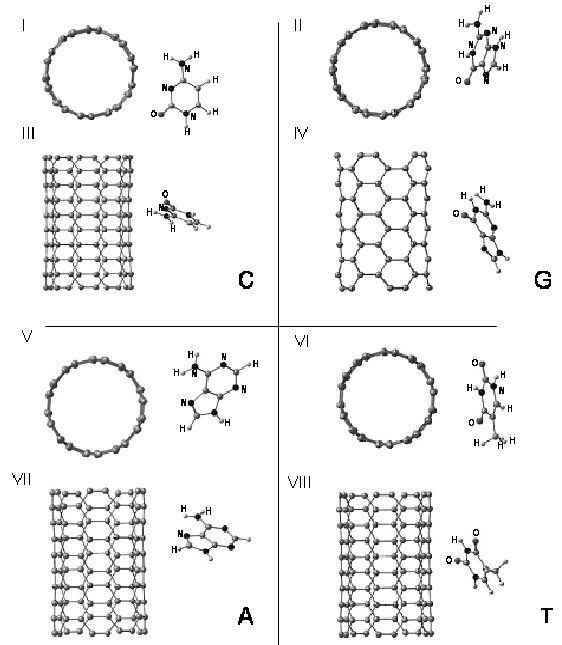}
\caption{I, II, V $\&$ VI show the cross-sectional view of optimized structure of C, G, A and T nucleobases bonded to nanotube. III, IV, VII $\&$ VIII show the lateral view of the corresponding optimized structures.}
\label{Figure 1}
\end{center}
\end{figure}

\section{Theoretical Calculations}

We consider a (5, 5) SWNT containing 5 unit cells in presence of different nucleobases (C, G, A, T) as input configuration for our quantum chemical calculations. The bases are kept parallel to the nanotube axis. During the optimization process, all the atoms are free to relax. The optimized structures of nucleobases, SWNT and combined system are obtained using Jaguar computational package \cite{jaguar}. The binding energy of nucleobases with SWNT is evaluated using ab-initio restricted open shell Hartree-Fock (ROHF). In all the calculations, we have used double-zeta basis set with polarization function (6-31g**). The optimized geometries for the combined systems are shown in Fig. 1. We notice that a particular orientation of the base with the nanotube, as shown in Fig. 1 is achieved irrespective of different input configurations. Hartree-Fock (HF) or density-functional theory (DFT) describes the charge transfer and chemical bonds between the molecules but they are inadequate to describe the van der Waals (vdW) interaction \cite{ortmann,williams}. It is known that the vdW is the main interaction between organic molecules and inert surfaces like graphite \cite{ortmann,sowerby}. Here we have calculated the vdW energy between nanotube and different bases using classical MSCFF force field \cite{brameld} as well as AMBER 7 \cite{amber}. The solvation energies \cite{tannor,tsang} were calculated using Poisson-Boltzman (PB) solver of Jaguar which takes into account the presence of continuum dielectric medium like water. The total binding energy is thus obtained by adding the HF, vdW and solvation energy. We show that the binding energy between the nanotube and DNA bases is mainly governed by vdW interaction, whereas the sequence of binding is influenced by the solvation energy.

\begin{figure}[tbp]
\begin{center}
\leavevmode
\includegraphics[width=0.425\textwidth]{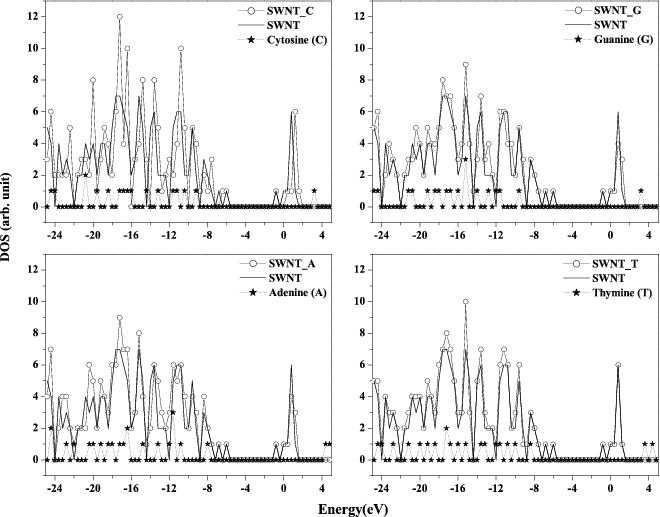}
\caption{The open circle with line shows the DOS for the combined system (SWNT\_nucleobase), whereas the solid line and stars with dashed line show the DOS for pristine nanotube and nucleobases, respectively.}
\label{Figure 2}
\end{center}
\end{figure}

\subsection{Ab-initio Calculations}

To investigate the binding energy of different bases with nanotube, we have calculated the energy of optimized structures of the nanotube, the bases and the combined system separately using Hartree-Fock (HF-6-31g**). The binding energy of a base is calculated by subtracting the energy of an isolated nanotube and isolated base from the energy of the combined system i.e E $_{binding}$ = E $_{adsorbed}$ $_{system}$ - (E $_{isolated}$ $_{nanotube}$ + E $_{isolated}$ $_{base}$). In this scenario the binding sequence is C $>$ G $>$ A $>$ T and the corresponding binding energies are -0.07, -0.04, -0.03 and -0.02 eV, respectively.

To investigate the origin of binding of nucleobases with nanotube we will examine the different energetic contributions, Mulliken charge analysis and electronic density of states (DOS). It is seen that there is no noticeable contribution from covalent or Coulombic energy to the attraction between nanotube and bases. The only noticeable contribution in the binding energy obtained from HF (6-31g**) comes from the exchange energy term. Nuclear repulsion (E $_{\sf{A}}$), total one electron term (E $_{\sf{E}}$) and total two electron term (E $_{\sf{I}}$) are the three components in binding energy obtained from HF (6-31g**), where the total two electron term consists of Coulomb (E $_{\sf{J}}$) and exchange energy (E $_{\sf{X}}$). The corresponding energies for cytosine bonded nanotube are E $_{\sf{A}}$ = 62595.74 eV, E $_{\sf{E}}$ = -125302.13 eV and E $_{\sf{I}}$ = 62706.32 eV with total energy E binding = E $_{\sf{A}}$ + E $_{\sf{E}}$ + E $_{\sf{I}}$ = -0.07 eV. As HF does not give the exchange energy separately, we have calculated the exchange energy with correlation term (E $_{\sf{XC}}$) for cytosine bonded nanotube using density functional theory (DFT) at the theoretical level of B3LYP/6-31g**. The corresponding energies are E $_{\sf{A}}$ = 68743.93 eV, E $_{\sf{E}}$ = -137595.49 eV, E $_{\sf{J}}$ = 68851.62 eV and E $_{\sf{XC}}$ = -0.13 eV. Therefore E $_{\sf{I}}$ = E $_{\sf{J}}$ + E $_{\sf{XC}}$ = 68851.49 eV and the total energy E binding = E $_{\sf{A}}$ + E $_{\sf{E}}$ + E $_{\sf{I}}$ = -0.07 eV, which is same as the binding energy obtained from the HF. The binding of cytosine with nanotube can be explained only if we consider the E $_{\sf{XC}}$ term (-0.13 eV), because the sum of 1st three components of energies (E $_{\sf{A}}$ + E $_{\sf{E}}$ + E $_{\sf{J}}$) is a positive contribution of 0.06 eV. The sum E $_{\sf{A}}$ + E $_{\sf{E}}$ + E $_{\sf{I}}$ energy would be positive if there is no strong covalent bond between nanotube and cytosine. The positive value arises from purely repulsive Pauli barrier. It will be explained in the next paragraph that indeed there is no noticeable covalent bond between the nanotube and nucleobases. Therefore, the binding energy calculated from HF suggests that the weak attraction between nanotube and nucleobases comes mainly from the exchange energy.

The DOS of pristine nanotube, isolated bases and the combined system are shown in Fig. 2 which clearly shows that there is no shift in the DOS of combined system with respect to the isolated SWNT and the DOS of combined system is the sum of DOS of pristine nanotube and the corresponding base. We note that the (5, 5) nanotube has an energy gap of 4 eV due to the finite length of 1.1 nm. The nearest distances in optimized structures (see Fig. 1) between the nanotube to different nucleobases are C-O ~3.2 $\sf{\AA}$ (cytosine and guanine), C-H ~ 3.3 $\sf{\AA}$ (adenine) and C-H ~3.4 $\sf{\AA}$ (thymine), which certainly exceed the sum of the covalent radii of C-O ~ 1.5 $\sf{\AA}$ and C-H ~ 1.1 $\sf{\AA}$. Therefore, the bond between the nanotube and the bases is unlikely to be covalent.

\begin{figure}[tbp]
\begin{center}
\leavevmode
\includegraphics[width=0.475\textwidth]{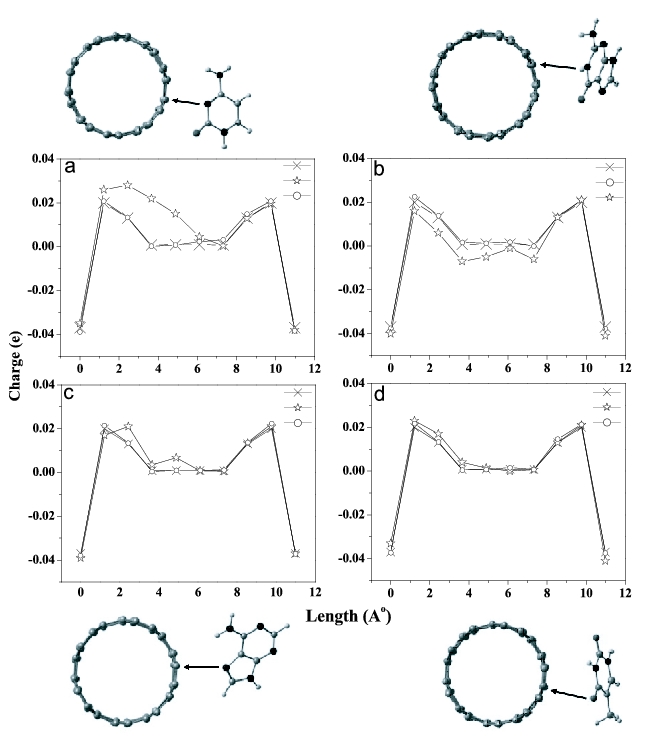}
\caption{The open star with line shows the charge distribution on individual carbon atoms along the nanotube axis near (a) the nitrogen of C, (b) hydrogen of G, (c) nitrogen of A, (d) oxygen of T. The cross with line shows the averaged charge on each carbon atom in a ring along the nanotube axis for neutral nanotube. The open circle with line shows the averaged charge on each carbon atom in a ring along the nanotube axis for nanotube bonded with different nucleobases.}
\label{Figure 3}
\end{center}
\end{figure}

	The adsorption energies in Hartree-Fock calculations are of similar magnitude as calculated for adenine adsorbed on Cu (110) \cite{preeuss}, where the mutual polarization and Coulombic interaction between the molecule and Cu substrate determines the binding energy. Guided by this, we have carried out the Mulliken charge population analysis to evaluate the charge transfer and charge distribution in the combined system. As shown in Fig. 3, the adsorption does cause a minor charge redistribution. However, there is no charge transfer between the nanotube and nucleobases. Out of all the four bases, the most pronounced case is that of cytosine bonded nanotube where the maximum charge redistribution of the order of $\Delta$q = + 0.01e occurs on a carbon atom of nanotube in close proximity to the nitrogen atom of the cytosine (see Fig. 3). This results an overall dipole moment in nanotube of $\left|\sf{P}\right|$ = 0.03 Debye (D = 3.33 $\times$ 10$^{-30}$ Coulomb meters). Since this charge redistribution is much smaller in magnitude compared to the adenine on copper \cite{preeuss}, we can exclude any noticeable contribution to binding energy due to mutual polarization between the nanotube and the bases. However, this mutual polarization can explain the origin for the specific orientation of the bases with respect to the nanotube in HF optimized structures. The optimized structures (Fig. 1) indicate that cytosine binds to nanotube surface via nitrogen atom of the pyrimidine ring with partial contribution from the oxygen of C=O group. Similarly for guanine, adenine and thymine, the binding occurs via either the nitrogen atom of the pyrimidine ring or the oxygen atom of the carbonyl group. As these nitrogen and oxygen atoms have more electron affinity, the resultant dipole moment of the pure bases is determined by these atoms.
	
\begin{figure}[tbp]
\begin{center}
\leavevmode
\includegraphics[width=0.475\textwidth]{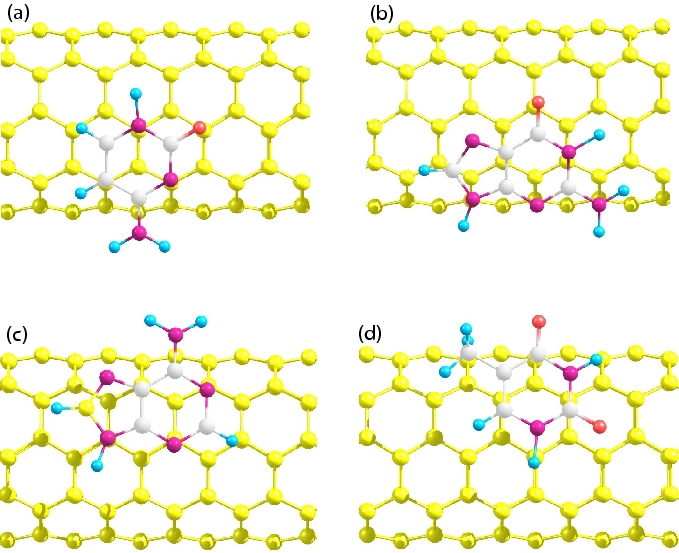}
\caption{(color online) Classically optimized geometry of nucleobases on nanotube: (a) cytosine, (b) guanine, (c) adenine and (d) thymine.}
\label{Figure 4}
\end{center}
\end{figure}	

\subsection{Classical Force Field Calculations}

Since the vdW interaction is the dominant interaction between organic molecules and nanotubes, we have carried out classical force field calculation to calculate the vdW energy between the nanotube and nucleobases. Here we have used MSCFF force field parameter \cite{brameld} between the different atoms of nanotube and nucleobases. The input configuration is the same as that in the quantum chemical calculations. From the optimized structures of combined systems, we note that all the nucleobases remain parallel to the nanotube at a distance of 3.25 $\sf{\AA}$ as shown in Fig. 4. The parallel configuration is obtained because of maximum vdW interaction between the substrate and the molecule ans the stacking arrangement (see Fig. 4) is similar to AB stacking of graphite layers but it deviates frob perfect AB stacking, which is more prominent in G, due to five and six-membered rings of bases and nanotube curvature effects. The vdW binding energies are calculated by subtracting the vdW energy of isolated systems from the energy of the adsorbed system and the corresponding energies are -0.54, -0.51, -0.47, -0.39 eV for G, A, T and C, respectively and the vdW energy calculated using AMBER 7 remains similar to that of MSCFF (see table 1). Now the resultant gas phase energies are obtained by adding the HF and vdW energies and the sequence becomes G $>$ A $>$ T $>$ C with energies -0.58, -0.54, -0.49, -0.46 eV, respectively (see table 1). It is clear that vdW energy has much more contribution to binding than the HF exchange energy. We note that the gas phase binding sequence of nucleobases with SWNT remains same as the binding sequence of nucleobases with graphite \cite{shi}.

	To compare the adsorption of nucleobases to nanotube and graphite, we have calculated the binding energy of all the bases with graphite. For adenine, the calculated binding energies using HF (6-31g**) and classical force field (vdW) are -1.50 kcal/mol (-0.06 eV) and -18.69 kcal/mol (-0.8 eV), respectively. Therefore, the total binding energy (HF+vdW) of adenine on graphite is -20.19 kcal/mol (-0.87 eV). These energies are very similar to the value of -0.07 eV calculated using DFT with Generalized Gradient Approximation (GGA) and the total energy of -1.07 eV (GGA + vdW) [26]. This comparison is very important as we calculate the the binding energy using HF and classical force field separately but it givesas accurate as DFT (GGA + vdW) calculation. The HF optimized structure for graphene-adenine combined system is shown in Fig. 5(a) and 5(b). We see that unlike nanotube adenine remains almost in parallel with graphene. The vdW interaction for graphite with other bases G, T and C bases are, respectively, -0.83, -0.75 and -0.7, eV. These values compare well with the reported values \cite{shi,gowtham} of graphite-base interaction using molecular dynamics and DFT calculation. We note that vdW interaction for the nanotube is smaller than the graphite due to curvature effects. The clasically optimized structures for different bases with graphene are shown in Fig. 5(c), (d), (e) and (f) for C, G, A and T respectively. Like in nanotube, $\pi$-orbitals of nucleobases and graphene minimize their overlap to lower the repulsion interaction giving rise to AB stacking like arrangement, which is more perfect than nanotube.
	
\begin{figure}[tbp]
\begin{center}
\leavevmode
\includegraphics[width=0.475\textwidth]{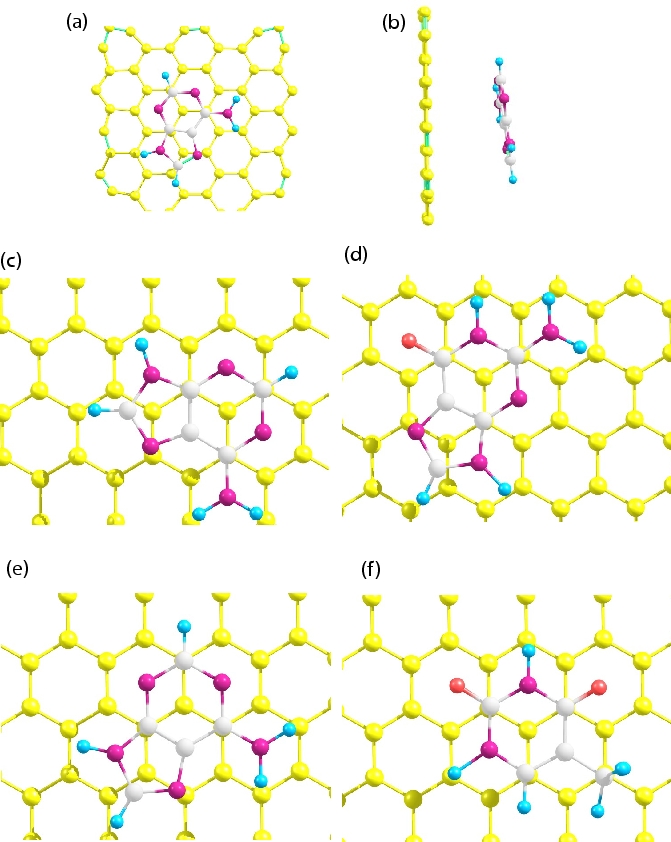}
\caption{(color online) (a) HF optimized structure for graphene-adenine combined system. (b) cros-sectional view. Classically optimized structures for (c) cytosine, (d) guanine, (e) adenine and (f) thymine.}
\label{Figure 4}
\end{center}
\end{figure}

\subsection{Solvation Energy Calculations}

The binding energies calculated so far are in gas phase condition. In real experimental situations, the nucleobases bind with the nanotubes in aqueous solution like water. To calculate the binding energy in solution phase we have carried out solvation energy calculations using Poisson-Boltzman (PB) \cite{tannor} solver of Jaguar package at the HF/6-31g** level, where the gas phase optimized structures (HF) are taken as input and the solvation energies (S.E) are calculated without optimizing the geometry further in presence of water. Here we have assigned the van der Waal radii for different atoms according to the reference \cite{lee}. Now the contribution of solvation energy to the binding energy is calculated by subtracting the solvation energy of isolated systems from the solvation energy of adsorbed system and the energies are -0.02, 0.01, 0.05 and 0.16 eV for T, G, A and C, respectively (see table 2). The total binding energy is obtained by adding the HF, vdW and the solvation energy. The relative binding for the monomeric bases is G $>$ T $>$ A $>$ C with energies -0.57, -0.51, -0.49 and -0.3 eV, respectively. 

\begin{table}[h!b!p!]
\caption{Theoretically calculated values of interaction energies of monomeric nucleobases with SWNT}
\begin{center}
\begin{tabular}{|c|c|c|c|c|c|c|}
\hline 
Nucleo&HF& vdW(eV)&HF+vdW&S.E&Total&vdW(eV)\\ 
base&eV&MSCFF&eV&eV&eV&AMBER\\
\hline 
Guanine & -0.04 & -0.54 & -0.58 & 0.01 & -0.57 &-0.6\\
\hline
Thymine & -0.02 & -0.47 & -0.49 & -0.02 & -0.51 &-0.5\\ 
\hline  
Adenine & -0.03 & -0.51 & -0.54 & 0.05 & -0.49 &-0.53\\ 
\hline
Cytosine & -0.07 & -0.39 & -0.46 & 0.16 & -0.3 &-0.45\\ 
\hline
\end{tabular}
\end{center} 
\label{table2}
\end{table}

\begin{figure*}[tbp]
\begin{center}
\leavevmode
\includegraphics[width=\textwidth]{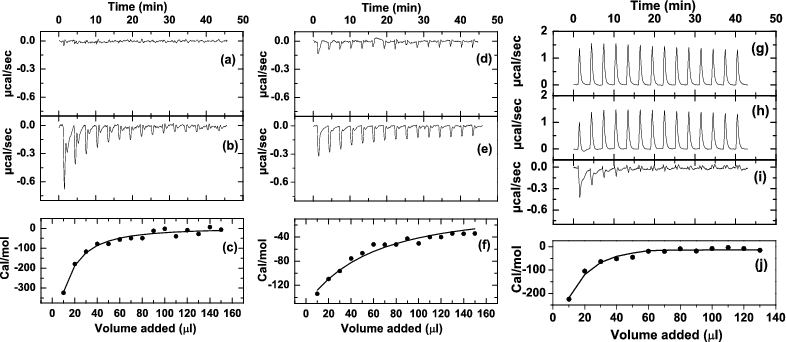}
\caption{The blank titration for T (a), C (d) and A (g). The raw ITC response for T (b), C (e) and A (h). (i) Exothermic reaction of adenine with SWNTs after subtracting the blank titration from the raw titration. (c), (f) and (j) show the integrated heat of reaction at each injection for T, C and A, respectively.}
\label{Figure 2}
\end{center}
\end{figure*}

\section{Experimental Details}

SWNT bundles were prepared by the arc discharge method followed by purification process, involving acid washing and high-temperature hydrogen treatment \cite{vivekchand}. The nanotubes are characterized by the thermogravimetric analysis, transmission electron microscopy (TEM), Raman spectroscopy [RS] and near infrared spectroscopy (NIR) \cite{vivekchand}. The average diameter of the nanotubes was about 1.4 nm, the length being a few microns. 100 $\mu$g of the SWNT sample was suspended in 1.6 ml double distilled water having a pH value of 6.7. The solution was sonicated for 5 hrs at a power level of 3W. After the sonication, the solution was kept for a day to sediment all the big bundles of SWNT. The supernatant solution was stable for more than 15 days. The nucleobases obtained from the Sigma Aldrich Chemicals were used as received. The binding of nucleobases with SWNTs was achieved by mixing of 300 $\mu$l of 10 mM nucleobases solution (prepared in double distilled water with pH of 6.7) with 1.6 ml of the above SWNT solution. We have chosen to suspend the nanotubes and the nucleobase in water instead of a buffer solution because the latter may influence the surface of the nanotubes. We could not perform the experiment with guanine in aqueous solution due to its insolubility in water. 
Calorimetric titrations were performed at 5 $^0$C with a Micro-Cal VP-ITC instrument. We checked the stabilization of the instrument by performing a water-water run, the heat change so obtained being less than 0.6 $\mu$Cal per 10 $\mu$l injection of water. Here a rotating stirrer-syringe (270 rpm) containing 300 $\mu$l of 10mM nucleobase solution injects in equal steps of 10 $\mu$l solution at 180-sec intervals into a cell containing 1.4 ml of SWNT solution until saturation is reached.

\section{Experimental Results}

The ITC response recorded during the titration of nucleobases with the SWNTs are shown in Fig. 6. Fig. 6a, 6d and 6g show the ITC data for the blank titration of 10 mM thymine, cytosine and adenine, respectively against double distilled water. Fig. 6b, 6e and 6h show the raw data of heat of reaction due to nucleobase-SWNT binding during each injection for thymine, cytosine and adenine, respectively. We can see that the blank titration of adenine (see Fig. 6g) with double distilled water is an endothermic reaction with higher magnitude of heat of reaction. The raw ITC response for adenine (see Fig. 6h) with SWNT solution also shows an endothermic response with different magnitude of heat of reaction. Therefore, the heat of reaction of adenine with SWNTs is obtained by subtracting the blank titration from the raw titration, as shown in Fig. 6i. The nature of the binding curve (Fig. 6b, 6e and 6i) arises due to lowering of accessible surface area of SWNT in successive injections as nucleobases bind with the SWNTs. The exothermic tail in the main curve is due to the dilution effect of nucleobases during the titration as there is no free surface of SWNTs available to bind. We have also seen that the heat of dilution of SWNT solution with water is less than 5 cal per 10 $\mu$l injection of water. The raw ITC of thymine (Fig. 6b) shows the appearance of second peak in each injection, which may be as a result of more available free surfaces due to possible removal of entanglement between the nanotube bundles. Fig. 6c, 6f and 6j correspond to the integrated heat change (enthalpy change in cal/mol) due to each injection of thymine, cytosine and adenine, respectively after correcting for heat of dilution (at each injection) and are plotted as a function of injected nucleobase volume. The experiments were repeated twice and show good reproducibility as seen by the relatively small standard errors. 

The strength of interaction of the nucleobases with SWNTs is evaluted from the exothermicity (integrated heat in Fig. 6c, 6f and 6j) of the first injection of nucleobases with SWNTs and the binding sequence is T (-320 cal/mol) $>$ A (-219 cal/mol) $>$ C (-134 cal/mol). Since the heat of reaction curve in Fig. 6 are not in sigmoidal nature, the absolute value of binding energy can not be calculated from Fig. 6. This is because of the limited solubility of the nanotubes as well as its bundle effect in aqueous solution.


\section{Conclusions}

Our results from HF combined with forcefield calculation indeed suggests that the binding of nucleobases with SWNTs is mainly due to weak vdW force and the gas phase binding sequence is G $>$ A $>$ T $>$ C like the binding sequence of nucleobases with graphene. We also show that compared to graphene, nanotube has lesser amount of binding affinity with nucleobases duo to curvature effects. We have shown that the solution phase binding sequence of nucleobases with nanotube is G $>$ T $>$ A $>$ C, which explains the earlier observations that oligonucleotides made of thymine bases are more effective in dispersing the SWNT in aqueous solution as compared to poly (A) and poly (C). We have shown by direct calorimetry that there is differential binding affinity in the interaction of nucleobases with SWNTs, the sequence being T $>$ A $>$ C, in agreement with the calculation.\\ 

\textbf{Acknowledgment}

We thank the Department of Science and Technology for financial support.

\end{document}